\documentclass[a4paper,10pt]{Swaluw}
\usepackage{graphicx}
\newcommand{\halfskip}{\vskip 0.5\baselineskip \noindent}
\newcommand{\be}{\halfskip \begin{equation}}
\newcommand{\ee}{\end{equation} \halfskip}

\newcommand{\bm}[1]{\mbox{\boldmath $#1$}}
\newcommand{\bdot}{\: \bm{\cdot} \:}

\begin{document}
\authorrunning{E. van der Swaluw}
\title{Interaction of a magnetized pulsar wind with its surroundings}
\subtitle{MHD simulations of Pulsar Wind Nebulae}
\author{E. van der Swaluw}
\institute{Dublin Institute for Advanced Studies, 5 Merrion Square, Dublin 2, 
Ireland
}
\offprints{E. van der Swaluw,
\email{swaluw@rijnh.nl}}
\date{}
\abstract{Magnetohydrodynamical simulations are presented of a magnetized
pulsar wind interacting directly with the interstellar medium, or, in the case
of a surrounding supernova remnant, with the associated freely expanding ejecta
of the progenitor star. In both cases the simulations show that the pulsar 
wind nebula will be elongated due to the dynamical influence of the toroidal magnetic fields, 
which confirm predictions from a semi-analytical model presented by Begelman \& Li. 
The simulations follow the expansion of the pulsar wind nebula when the latter is
bounded by a strong shock and show that the expansion can be modeled with a standard power-law
expansion rate. By performing different simulations with different magnetization parameters, 
I show that the latter weakly correlates with the elongation of the pulsar wind nebula. 
The results from the simulations are applied to determine the nature of the expansion
rate of the pulsar wind nebula 3C58. It is shown that there is both observational and
theoretical evidence which supports the scenario in which the pulsar wind nebula 3C58 has 
caught up with the reverse shock of the associated (but undetected) supernova
remnant.
}
\maketitle

\section{Introduction}

After the explosion of a massive star as a supernova, the progenitor star 
collapsed into a compact object, which can be a neutron star or a black hole. If 
the neutron star manifests itself as a rapidly rotating
pulsar, the spin-down energy of the pulsar is converted into a relativistic
pulsar wind (Rees \& Gunn 1974), which blows a bubble or pulsar wind nebula (PWN) 
into the surrounding medium. The observational counterpart of a PWN is a plerion,
which is characterized by filled-center diffuse emission (Weiler \& Panagia 1978).
For young plerionic supernova remnants, the central region contains the active young 
pulsar.

The supernova explosion itself drives a blastwave into the interstellar medium (ISM),
which results in an expanding shell of hot (shocked) material. The observational
counterpart is a shell type supernova remnant, which is characterized by extended
emission from the roughly spherical shell bounding the supernova remnant (SNR).
Due to the deceleration of the expanding SNR by the surrounding ISM, a reverse shock
forms (McKee 1974), which will propagate into the interior of the SNR when the amount 
of swept-up material becomes comparable with the ejected mass of the progenitor star. 
In the case of a SNR which contains both, an expanding shell of hot shocked material 
and a centered PWN (which is the observational counterpart of a composite remnant), the 
interaction between both components becomes significant when the reverse shock hits 
the PWN (Reynolds \& Chevalier 1984; van der Swaluw et al. 2001; Blondin et al. 2001).
After the passage of the reverse shock the PWN will make a transition from a
supersonic to a subsonic expansion (van der Swaluw et al. 2001). 
The PWN will be deformed into a bow shock nebula when the motion of the pulsar becomes
supersonic as the pulsar approaches the shell of the SNR (van der Swaluw et al. 2003).

In this paper I will focus on young PWNe which are expanding supersonically, where the
surrounding medium can be either the ISM or the freely expanding ejecta of the
progenitor star (i.e. the stage of the PWN before it encounters the reverse shock).
For both scenarios there exists observational evidence. In the case of SNR G11.2-0.3
a large PWN radius with respect to the shell of the SNR is detected at radio frequencies
(Tam et al. 2002), making the system a probable candidate for a PWN which has not encountered 
the reverse shock yet. In the case of the Crab Nebula there is indirect 
evidence for the 
presence of a shell containing shocked material and fast moving ejecta around the pulsar 
wind nebula (Sankrit \& Hester 1997). There are some counter-examples, naked plerions like
G74.9+1.2 and G63.7+1.1, where there is no evidence for any shell or fast-moving ejecta 
around the plerionic component (see e.g. Wallace et al. 1997a\& b). This suggests that 
the pulsar wind in these systems is interacting directly with the surrounding ISM.  

A common feature of most plerions is a morphology with a significant elongation. This
elongation is most strongly apparent at radio frequencies, where the complete bubble 
can be observed because of the long lifetime of radio electrons with respect to the 
age of the PWN itself. Begelman \& Li (1992) (hereafter B\&L) explain the 
elongation of the Crab Nebula by considering an axially symmetric model, where
the elongation results from the dynamical influence of the toroidal magnetic 
fields in the pulsar wind bubble. Their model is a steady-state model, in contrast with 
recent time-dependent models of PWNe evolving inside SNRs as discussed by van der Swaluw 
et al. (2001) and Blondin et al. (2001). However, these authors performed hydrodynamical 
simulations, and thus could not consider the dynamical influence of the
magnetic fields on the PWN evolution.  

In this paper I consider the time-dependent evolution of a supersonically expanding 
PWN, driven by a magnetized pulsar wind. The evolution of the PWN is 
modeled by performing magnetohydrodynamical (MHD) simulations using the Versatile 
Advection Code \footnote{See http://www.phys.uu.nl/\~{}toth/} (VAC). I confirm the earlier 
analytical results from B\&L, which state that the elongated morphology of the Crab 
Nebula (and other plerions) can be explained by the dynamical influence of the toroidal 
magnetic fields in the pulsar wind bubble.

I trace the position of the shock during the evolution of the PWN and show
that although the magnetic fields are dynamically important, the expansion rate
can still be modeled by a power-law expansion. I show
that during the overall expansion the elongation will increase at a very slow rate,
and it is at the very early stage of the PWN that its shape will be significantly 
elongated. Furthermore I will discuss the importance of the effects of the magnetization
parameter $\sigma$, i.e. the standard ratio of the Poynting flux to the plasma kinetic 
energy flux, in the pulsar wind on the evolution of the PWN. Finally I will discuss the 
implications of the current simulations on the interpretation of the evolutionary status 
of the PWN 3C58.

\section{Evolution of a pulsar wind bubble}
\subsection{The hydrodynamical limit}
I consider the evolution of a supersonically expanding PWN. This type of expansion 
occurs when the pulsar wind is directly interacting with the ISM or with the freely 
expanding ejecta of a young SNR. In this section I consider the hydrodynamical limit, 
i.e. no magnetic fields are involved, which enables one to use simple power-law expansion 
rates for the radius of the PWN ($R_{\rm pwn}\propto t^{\alpha}$).

In the case of a pulsar wind with a constant luminosity $L_0$, interacting with
the ISM with a uniform density $\rho_0$, the eqation for the radius of the PWN
is similar to the equation for stellar wind bubbles (Castor et al. 1975), i.e.
\begin{equation}
  R_{\rm pwn} = \bar C\left({L_0\over\rho_0}\right)^{1/5}t^{3/5},
\end{equation}
where $\bar C$ is of the order of unity. 
In the case of a PWN expanding in the freely expanding ejecta of a SNR, the 
equation for the radius of the PWN was derived by van der Swaluw et al. (2001):
\be
\label{Anpwn}
        R_{\rm pwn} = C\left({L_0t\over E_0}\right)^{1/5}V_0t\propto t^{6/5},
\ee
here $L_0$ is the spindown luminosity, $E_0$ is the explosion energy of the SNR,
$C$ is of the order of unity and $V_0$ is defined by:
\be
\label{MechSNR}
        E_0 = \mbox{$\frac{3}{10}$} \: M_{\rm ej} V_{0}^{2} \; ,
\ee 
here $M_{\rm ej}$ denotes the mass of the ejecta.

The ratio ${\cal R}$ between the radius of a PWN expanding into the ISM
(equation 1) with respect to a PWN expanding into freely expanding ejecta (equation 2) is 
of the order of unity for supersonically expanding PWNe (using $\bar C = C =1$):
\[
        {\cal R} \simeq 2.36 \left({n_0\over E_{51}}\right)^{1/5} V_{10\; 000} t_{\rm kyr}^{3/5} \; , 
\]
here $V_{10\; 000}$ is expressed in units of $10\; 000$ km/sec, $E_{51}$ is the eplosion energy
in units of $10^{51}$ ergs, $n_0$ is the hydrogen number density of the ISM in units of
cm$^{-3}$ and $t_{\rm kyr}$ is the age expressed in units of $1\; 000$ years.
The simulations indeed show an enhanced expansion rate for the pulsar wind expanding into the ISM for the
considered ages ($t_{\rm kyr}\simeq 1)$. This can be explained by the fact that the pulsar wind 
embedded in the freely expanding ejecta will start to blow its bubble into the very dense medium of the
ejected material from the supernova explosion. However, as the ejecta is freely expanding the 
density will decrease as a function of time ($\rho_{\rm ej}\propto t^{-3}$) which will {\it accelerate}
the overall expansion of the pulsar wind bubble (e.g. Chevalier (1982); van der Swaluw et al.
(2001)). This behaviour is reflected in the equations above by inserting $\rho_0\sim t^{-3}$ into equation 
(1) which yields the power-law behaviour of equation (2): the expansion of the pulsar wind bubble starts 
at a slow rate due to the dense medium but accelerates to high expansion rates as the density of the ejecta
is uniformly decreasing. 

I will use the MHD simulations to
investigate the expansion rate of a magnetized PWN, and compare the expansion
rate with the above power-law expansion rates.

\subsection{A magnetized quasi-static pulsar wind nebula}
A magnetized pulsar wind is driven by the spin-down luminosity of a rapidly rotating 
pulsar. The pulsar wind has a high Lorentz factor (in the case of the Crab 
$\Gamma_{\rm w} \ge 10^{3}$ (Gallant et al. 2002)) and is terminated by a 
strong MHD shock, where the wind pressure and the confining ram pressure balance.
At the position of the wind termination shock $R_{\rm ts}$ one can write the luminosity 
of the pulsar wind as a combination of particle energy and magnetic energy (Kennel 
\& Coroniti 1984):   
\begin{equation}
        L = 4\pi \Gamma_{\rm w}^{2} n_{\rm w} R_{\rm ts}^{2}  
        m c^{3}(1+\sigma) \; ,
\end{equation}
where $n_{\rm w}$ is the proper density in the wind, $m$ the mean mass
per particle and $\sigma$ is the ratio of magnetic energy to particle 
energy, i.e.:
\be
\sigma = {B^2\over 4\pi n_{\rm w} \Gamma_{\rm w} mc^2},
\ee
where $B$ is the magnetic field component, which I will assume to be 
purely toroidal:
\begin{equation}
   {\bm B}=B(r,z){\bm {\hat\phi}},
\end{equation}
using an axially symmetric configuration (r, $\phi$, z).

The shock heats the incoming flow and yields a high sound speed, 
$c_{\rm s}\sim c/\sqrt{3}$ downstream of the MHD shock (Kennel \& Coroniti 1984). 

B\&L described the interior of a PWN by considering a steady-state solution 
($\partial / \partial t =0$) for the ideal MHD equations in an axially symmetric 
configuration. 
Due to the high sound speed in the PWN, pressure imbalances are quickly smoothed
out, therefore one can write the momentum equation as:
\be
{\bm\nabla}\left(P + {B^2\over 8\pi}\right)\; + \; {B^2\over 4\pi r}{\bm{\hat
r}} = 0,
\ee
i.e.
\begin{equation}
{\partial\over\partial z}\left(P+{B^2\over 8\pi}\right) = 0,
\end{equation}
and
\begin{equation}
{\partial\over\partial r}\left(P+{B^2\over 8\pi}\right) = -{B^2\over 4\pi r}.
\end{equation}
From these two equations one can immediately observe that the total pressure
along the symmetry axis (the z-axis) will be constant (equation 8), whereas the 
total pressure along the radial axis will decrease; the pressure gradient in the radial 
direction balances the magnetic pinching force (the RHS of equation 9), therefore 
the total pressure 
along the r-axis will decrease (equation 9). B\&L show that the resulting difference 
along the PWN shock in pressure between the interior and the exterior of the bubble will 
result in an elongation of the PWN, as is observed in the Crab Nebula today. 

\section{MHD simulations}

\subsection{MHD calculations of a pulsar wind}
The difference between the model presented in this paper and the model of B\&L is
that I integrate the complete set of MHD equations in the non-relativistic 
limit with a continuous (thermal and magnetic) energy and mass input in a region centered
around the position of the pulsar to model the pulsar wind. This yields the complete 
evolution of a PWN, for which one can trace the expansion, elongation rate and the internal 
structure, without being restricted to a quasi-static model as in B\&L.
 
I use the Versatile Advection Code (VAC; T\'oth 1996) to integrate the 
equations of ideal MHD:
\begin{eqnarray}
   & &  {\partial\rho\over\partial t} + 
   \nabla\bdot\left(\rho\bm{V}\right) 
   = 0  \; ; \\
   & & \nonumber \\   
   & &  {\partial (\rho \bm{V}) \over \partial t} 
   + \bm{\nabla} \:
   \bdot\left( \rho {\bm V}\otimes{\bm V}  + P \: \bm{I} -
   {{\bm B}\otimes{\bm B}\over 8\pi} \right) = 0 \; ;
   \\  
   & & \nonumber \\
   & &  {\partial\over\partial t}\left( {1\over 2}\rho V^2 +\rho e \right) 
   + \bm{\nabla} \bdot \left( \rho \bm{V} \: ({1 \over 2}V^2 +h -
   {B^2\over 8\pi})\: \right) = 0,
   \\
   & & \nonumber \\
   & & {\partial\bm{B}\over\partial t} + \bm{\nabla} \times 
       (\bm{V} \times \bm{B}) = 0,  
\end{eqnarray}
here $h$ is the enthalpy, 
\[
        h \equiv e +{P\over\rho} \; ,
\]       
$\rho$ is the mass density, $\bm V$ is the velocity, $B$ is the magnetic field
strength, $e$ is the internal energy density and $P$ is the thermal pressure.

\begin{figure}
\resizebox{\hsize}{!}
{\includegraphics[scale=0.7,angle=-90]{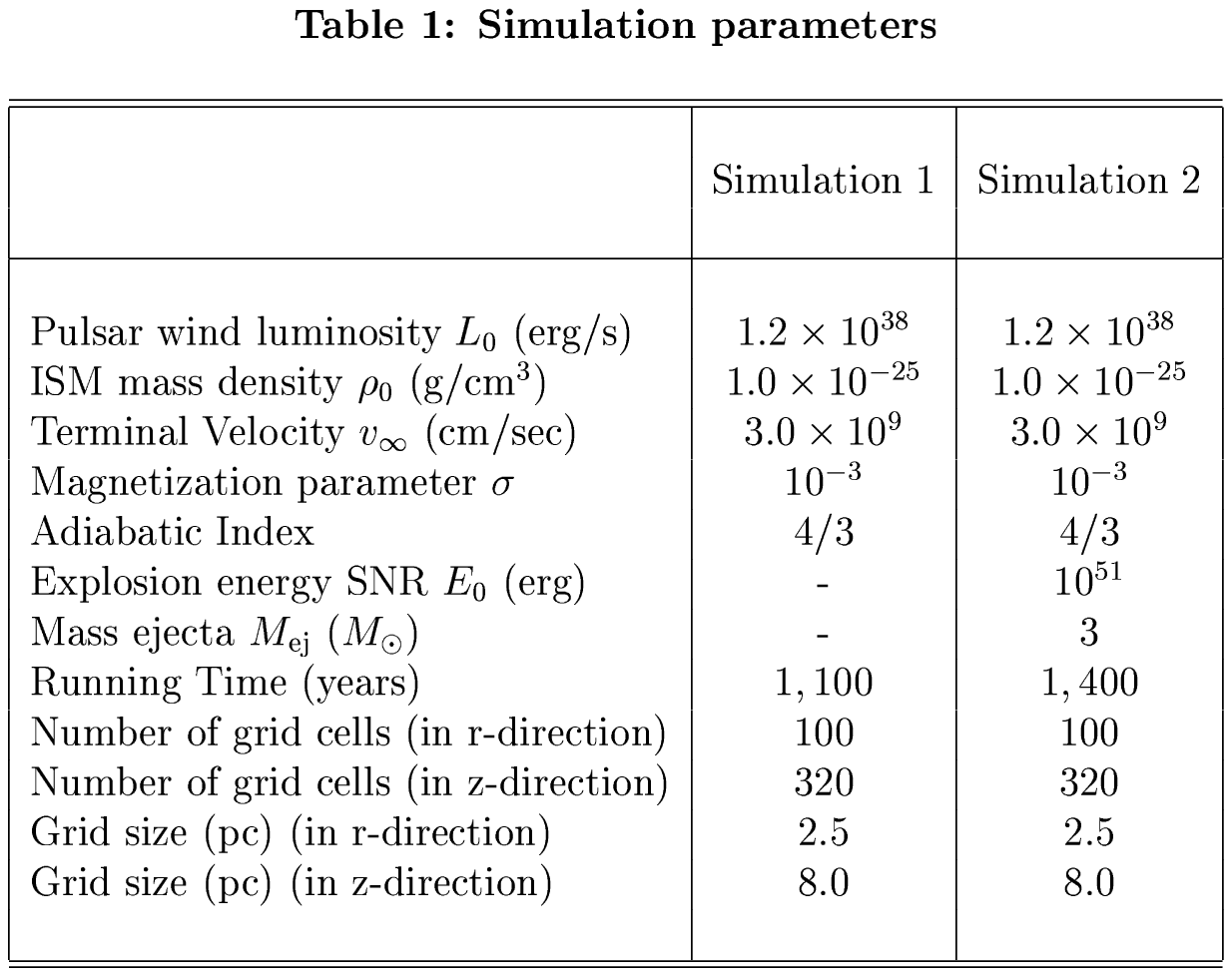}}
\end{figure}

I solve the MHD equations on a uniform grid in an axially symmetric configuration 
$(r,\phi ,z)$ in 2.5D. 
In a 2.5D simulation all the flow variables ${\cal F}_{\rm i}$ (where i=r,$\phi$,z) are 
solved as a function of $r$ and $z$, which includes the dynamical influence of the flow 
variables in the $\phi$ direction, but the flow variables in this direction are constant 
at a given $r$ and $z$, i.e. ${\cal F}_{\phi}(r,\phi ,z) ={\cal F}_{\phi}(r,z)$. I use a 
Total-Variation-Diminishing Lax-Friedrich scheme (T\'oth \& Odstr\v{c}il 1996) to solve the
above MHD equations in conservative form. The boundary conditions are continuous everywhere
except at the axis of symmetry where (a)symmetric boundary conditions were used for 
(momentum along the symmetry axis) density, total energy and the momentum along the r-axis.

\begin{figure}
\resizebox{\hsize}{!}
{\includegraphics{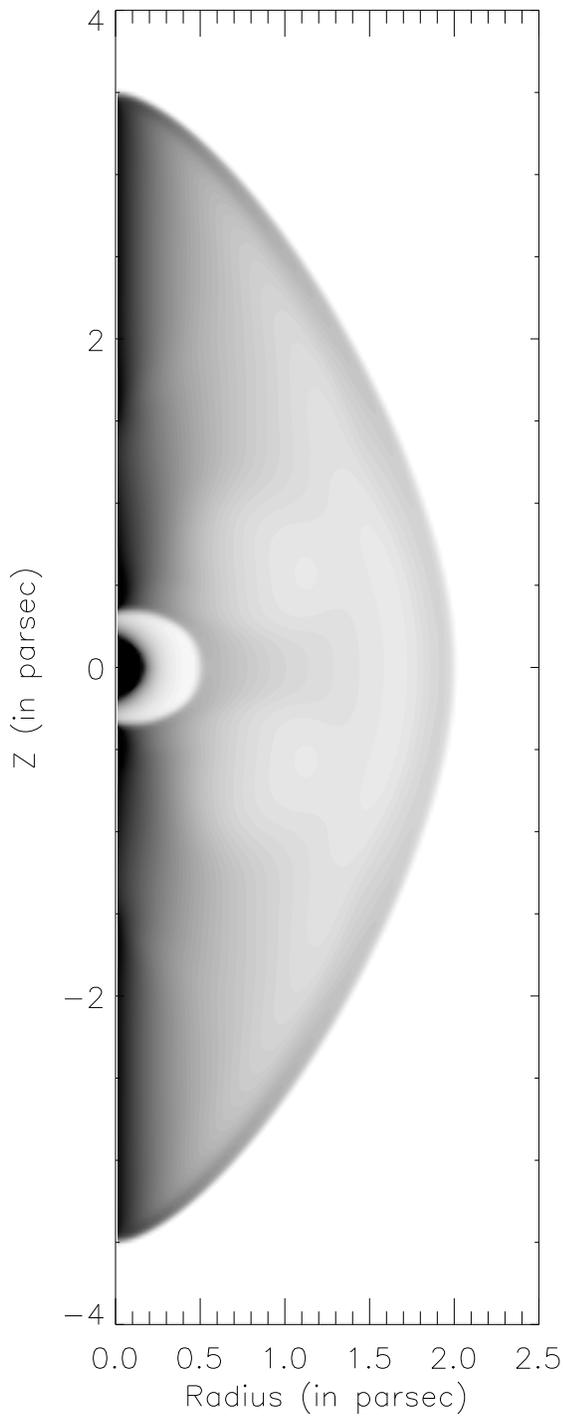}}
\caption{Gray-scale representation of the total pressure distribution of an elongated
PWN.}
\end{figure}

\begin{figure}
\resizebox{\hsize}{!}
{\includegraphics{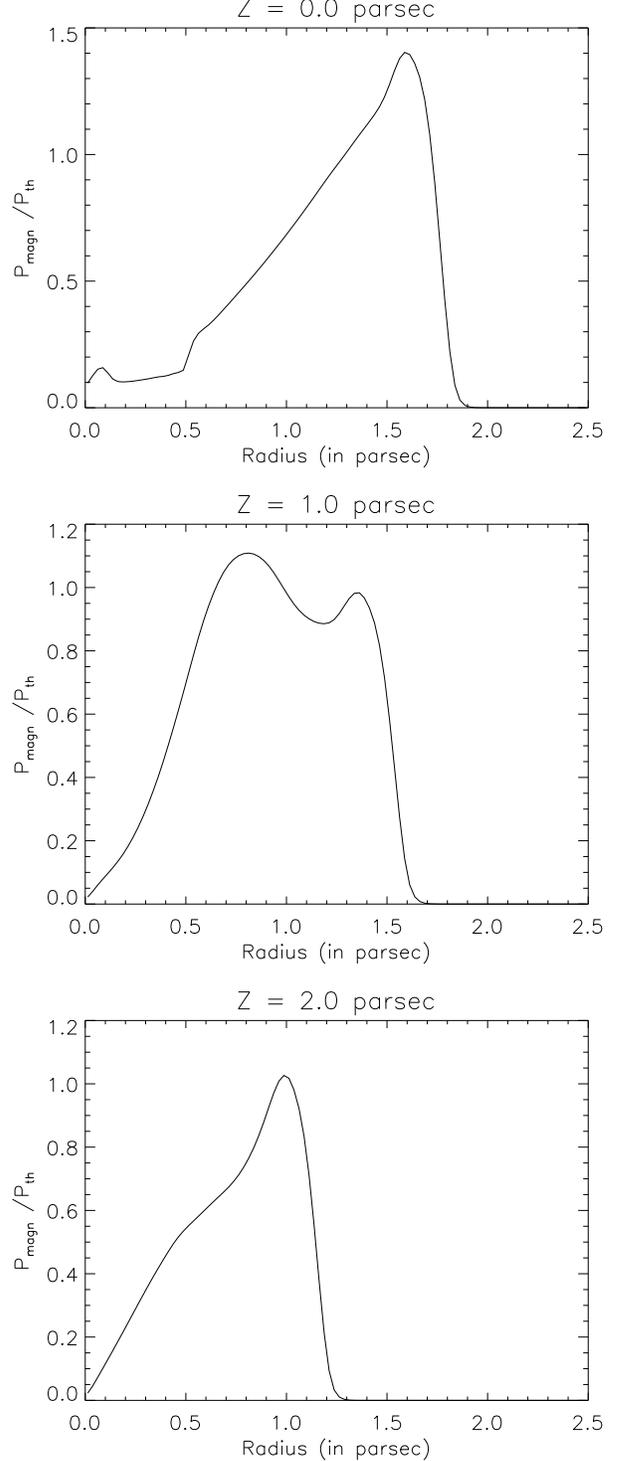}}
\caption{Ratio of magnetic pressure $P_{\rm magn}$, to the thermal pressure $P_{\rm th}$
throughout the PWN. A cut has been taken along the r-axis for different values of $Z$.}
\end{figure}

\begin{figure}
\resizebox{\hsize}{!}
{\includegraphics{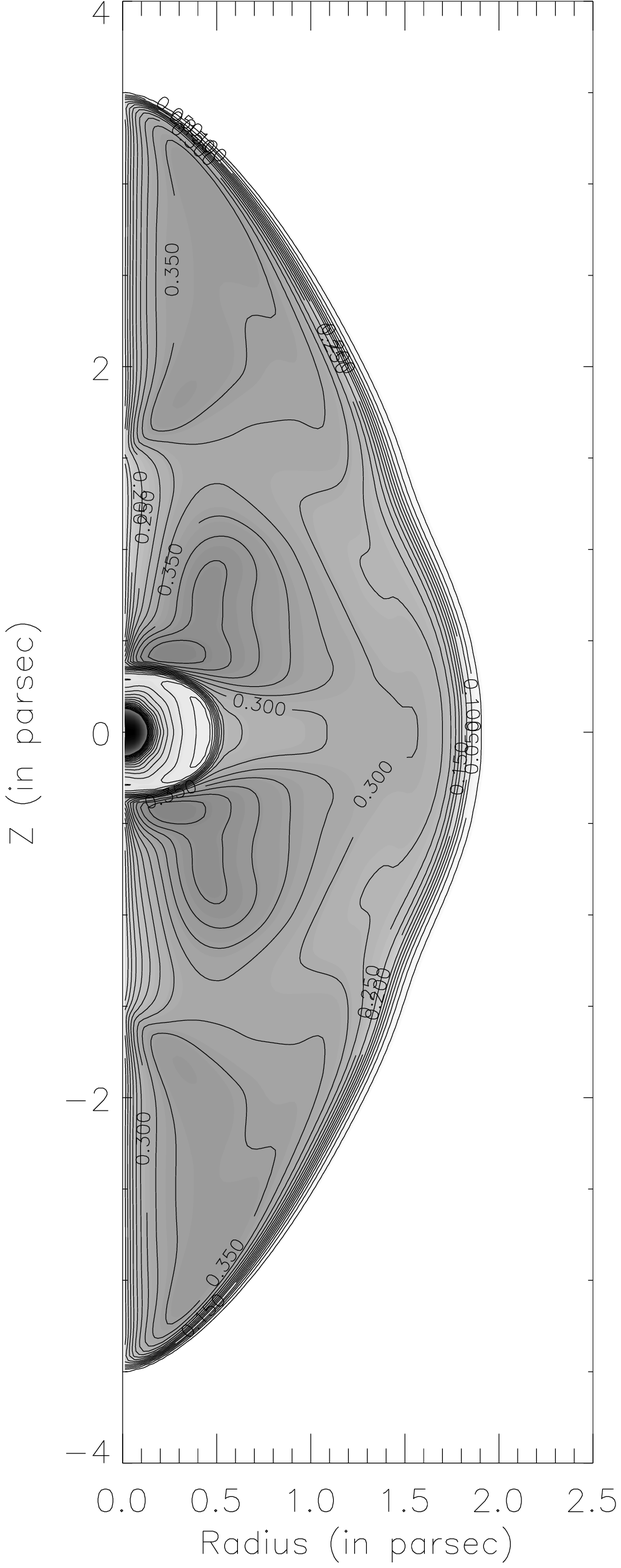}}
\caption{Gray-scale representation of the toroidal magnetic field distribution,
where the labels in the plot are in units of $10^{-4}$ Gauss.}
\end{figure}

\begin{figure}
\resizebox{\hsize}{!}
{\includegraphics{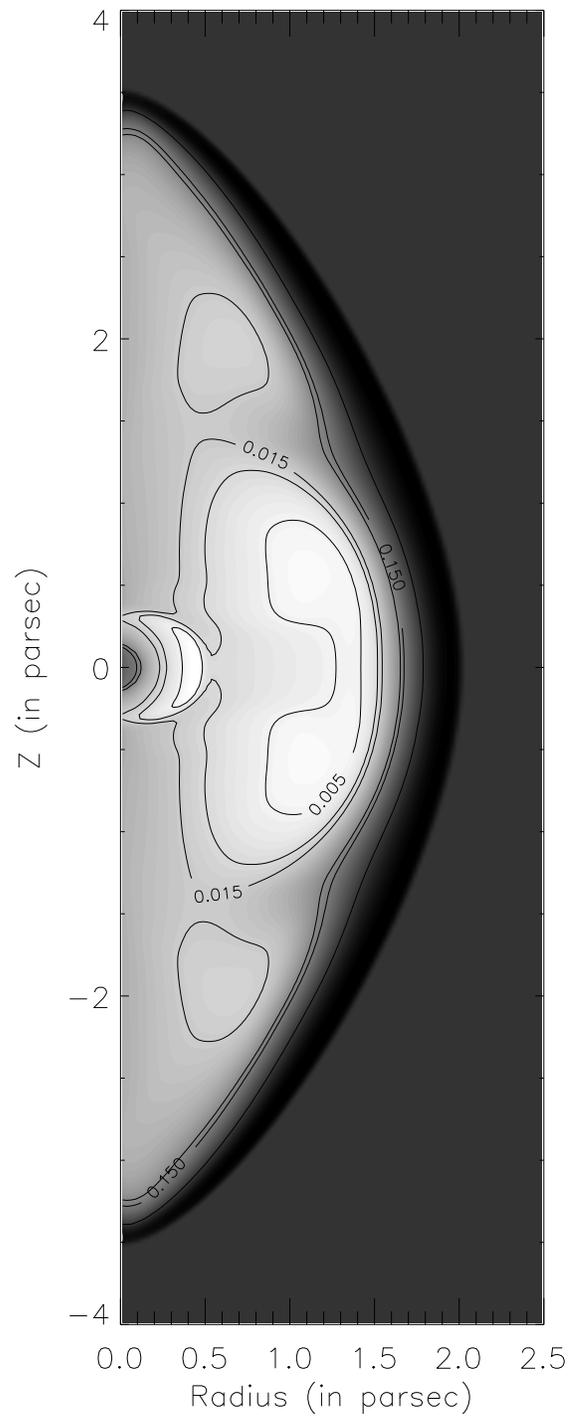}}
\caption{Logarithmic gray-scale representation of the density distribution,
where the labels in the plot are in units of $10^{-25}$ g/cm$^3$.}
\end{figure}


\subsection{Initial conditions of the pulsar wind}

I simulate a magnetized pulsar wind by depositing mass $\dot M$ and energy $L$ continuously
in a few grid cells concentrated around the position of the pulsar. The terminal velocity 
of the pulsar wind is determined from these two parameters, i.e. 
$v_{\infty}=\sqrt{2L/\dot M_{\rm pwn}}$. In all simulations performed in this paper the
terminal velocity has a value much larger then all the other velocities of interest.

I want to include the dynamical effect of the magnetic fields in the pulsar wind 
bubble, therefore part of the total luminosity $L_{\rm tot}$, is released as magnetic 
energy $L_{\rm magn}$, where the magnetization parameter $\sigma$, enters the equations:
\begin{equation}
L_{\rm magn} = \sigma L_{\rm tot} = B^2/2.
\end{equation}
Since we are considering a PWN which is much larger than the size $R_{\rm lc}$
of the light-cylinder of the pulsar involved, we can safely neglect poloidal magnetic
fields, therefore I will consider a purely toroidal magnetic field component.
Simple estimates (c.f. Contopoulos \& Kazanas, 2002) give a scaling law for the ratio of
the toroidal $B_{\phi}$, and poloidal $B_{\rm p}$ magnetic field strength,
\begin{equation}
B_{\phi}/B_{\rm p}\sim R/R_{\rm lc} \gg 1,
\end{equation}
which makes the above assumption a valid one.

Furthermore the current calculations do not solve the problem of transforming a high 
$\sigma$-wind into a low $\sigma$-wind (see e.g. Lyubarsky \& Kirk 2001), the simulation 
already starts with a low value for $\sigma$ as is appropriate at the MHD wind termination 
shock (Kennel \& Coroniti 1984). 

\subsection{A PWN in an uniform ISM}

The simulation of a magnetized pulsar wind interacting with a homogeneous ISM
has been performed with parameters as denoted in table 1 (simulation 1). The MHD 
code evolves till a corresponding age of $t=1,100$ years. Below I will discuss the 
profiles of the different flow variables of the system at the end of the simulation.

\begin{figure}
\resizebox{\hsize}{!}
{\includegraphics{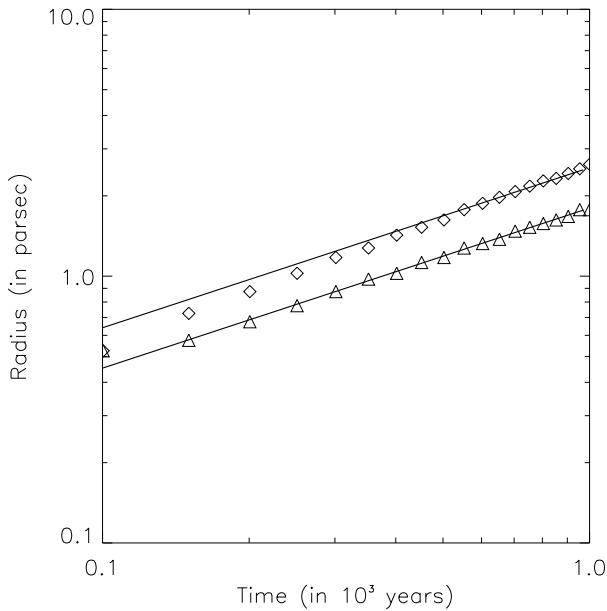}}
\caption{Distance between the minimum (open triangles) and maximum (open squares) 
distance between the shock and the position of the pulsar (denoted in the text by
$R^-$ and $R^+$ respectively). The solid lines are curves which follow a 
$R_{\rm pwn}\propto t^{3/5}$ law.}
\end{figure}

\begin{figure}
\resizebox{\hsize}{!}
{\includegraphics{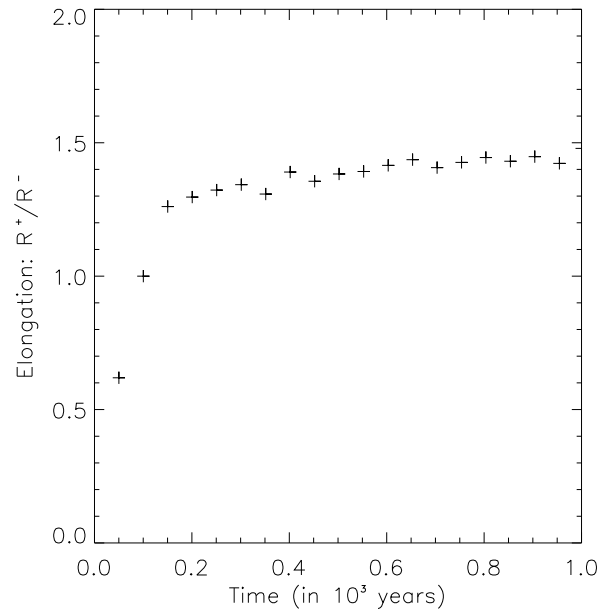}}
\caption{Ratio of $R^+$ to $R^-$, i.e. the elongation of the pulsar wind
nebula due to the presence of toroidal magnetic fields.}
\end{figure}

Figure 1 shows the total pressure (= thermal pressure {\it plus} magnetic pressure) 
profile. One can observe the presence of a pressure gradient along the r-axis, which 
balances the magnetic pinching force as discussed in the previous section. 

Figure 2 shows the ratio of the magnetic pressure  and the thermal pressure where a 
cut has been taken along the r-axis at different positions on the z-axis. The model
from B\&L gives a ratio of magnetic to gas pressure which is monotonically increasing
as a function of the cylindrical radius $r$ and is independent of the $z$ coordinate.
The model presented here allows for local maxima, which is also reflected in the
magnetic field distribution shown in figure 3. 


The density profile in figure 4 clearly shows the elongation of the PWN, where 
the ratio between the minimum and maximum distance between the shock bounding the 
PWN and the position of the pulsar as a function of time (denoted by $R^-$ and $R^+$ 
respectively), equals $\sim 1.5$. This results confirm the elongation of a PWN by the 
dynamical influence of toroidal magnetic fields inside the pulsar wind bubble as 
predicted by the steady-state model of B\&L. 

Figure 5 shows the evolution of the minimum and maximum distance between the
shock bounding the PWN and the position of the pulsar as a function of time 
($R^-$ and $R^+$). It is shown that the expansion rate for 
these shock positions still follow a simple power-law expansion rate, i.e. 
$R^{+/-}\propto t^{\alpha}$ with $\alpha =3/5$ as is appropriate for the
expansion law of a purely hydrodynamical (stellar) wind interacting with
a uniform ISM. This indicates that although the magnetic field influences the 
dynamics of the PWN, the evolution of a PWN driven by a magnetized pulsar wind 
can still be modeled by a standard power-law expansion rate.

Figure 6 denotes the ratio between $R^+$ and $R^-$, from which one can observe
that a significant part of the elongation of PWNe takes place in early stages 
($\sim$ 100 years) of its evolution. At later stages the elongation saturates at 
a very slow rate. In the last subsection I will investigate the dependency of the 
elongation on the magnetization parameter $\sigma$.

\subsection{A PWN in freely expanding ejecta}

In order to investigate the structures observed in the simulations presented
above are dependent on the environment in which the PWN expands, I performed a 
MHD simulation where a magnetized pulsar wind interacts with the freely expanding
ejecta of a supernova remnant. The parameters of the simulation are denoted in
table 1 (simulation 2). The freely expanding ejecta are initialised by impulsively
releasing the mechanical energy and the ejected mass corresponding to a supernova 
explosion in the first few grid cells located around the position of the pulsar. Due to 
the larger scales of the expanding SNR with respect to the PWN, both the reverse shock 
and the forward shock will move off the grid. The freely expanding ejecta however will 
be represented correctly by using continuous boundary conditions, which means that the 
gradient is kept zero by copying the values of the flow variables from the edge of the 
grid into the ghost cells located around the grid. Using the above initialisation of the
freely expanding ejecta of the SNR plus simultaneously simulating a pulsar wind as
described above one can investigate the dynamics of the same pulsar wind as performed 
in simulation 1, but interacting with a different environment. 

At the end of the simulation, at a  corresponding age of $1,400$ years, the
PWN has again been elongated along the symmetry axis. The overall structure of the 
different flow variables are similar to the ones discussed in the section above 
when the pulsar wind was interacting with a uniform ISM. As an illustration figure 7 
depicts the gray-scale representation of the magnetic field distribution which has the 
same overall structure as depicted in figure 3 for the first simulation.

From the above results I conclude that a magnetized pulsar wind in either a uniform
ISM or in freely expanding ejecta bounded by an expanding SNR, will yield an elongated
PWN with a similar distribution of the several flow variables. This seems to imply that
the shape and internal dynamics of a PWN is more determined by the central engine of the
system then the surrounding medium in which it expands as long as the PWN is bounded by
a shock.

\begin{figure}
\resizebox{\hsize}{!}
{\includegraphics{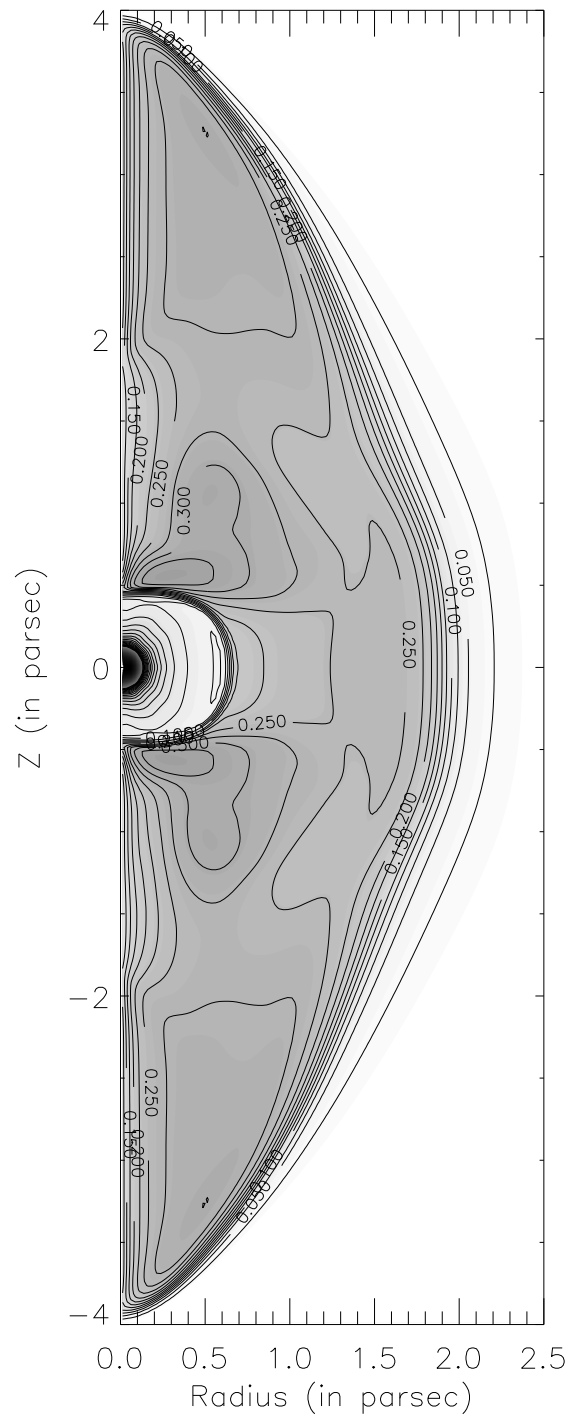}}
\caption{Gray-scale representation of the toroidal magnetic field distribution,
where the labels in the plot are in units of $10^{-4}$ Gauss.}
\end{figure}


\subsection{Elongation of pulsar wind nebulae}

The magnetization parameter determines the elongation rate of a supersonically 
expanding PWN. In order to quantitatively determine the influence of this 
parameter I performed several simulations for a magnetized pulsar wind with parameters
as in table 1 (simulation 1) but with different values for the magnetization parameter. 
All values are in the range of earlier models of PWNe (Kennel\& Coroniti 1984; 
Emmering \& Chevalier 1987; B\& L). Figure 8 summarises the results, which shows the 
elongation of the PWN as a function of time for the different MHD simulations. One can 
observe a weak dependency of the elongation on the magnetization parameter; an increase 
of roughly $\sim 1.44$ in elongation for an increase of 10 in the magnetization parameter. 
Furthermore, one can observe in all cases that the elongation saturates at late times in the
evolution of the PWN.

\begin{figure}
\resizebox{\hsize}{!}
{\includegraphics{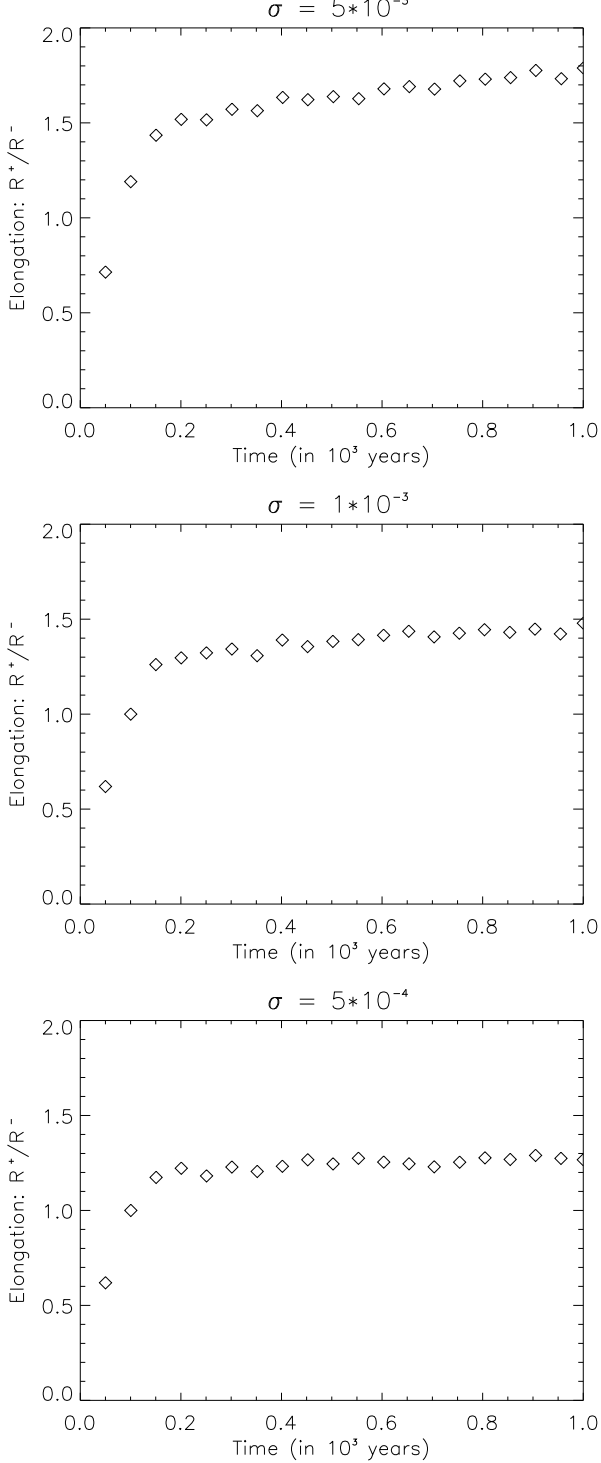}}
\caption{Elongation of a PWN driven by the same pulsar wind (parameters are denoted
in table 1 (simulation 1)) but with different values for the magnetization parameter 
$\sigma$.}
\end{figure}


\section{SNR 3C58, a crushed PWN?}

A beautiful example of a plerionic SNR which fits the morphology of the MHD
results as discussed above is 3C58. The shape of this remnant shows a
strong elongation at radio frequencies (see for example Bietenholz et al. 2001). 
Recently the pulsar of this remnant has been detected, which places the active pulsar 
in the center of the remnant (Murray et al. 2002). This detection makes 3C58 together 
with the Crab Nebula an excellent example of an elongated PWN blown into its surrounding 
medium by an active pulsar wind with the associated pulsar {\it detected}.  I have shown in 
this paper that although the magnetic fields in the bubble are dynamically important and 
necessary to explain the elongation, the expansion of the PWN can 
still be modeled with a power-law expansion rate. This means that independent of the 
density of the surrounding medium, the expansion velocity of the PWN will roughly equal
$V_{\rm exp}=\alpha R_{\rm pwn}/t_{\rm age}$, where $\alpha = 6/5$ when expansion takes 
place in the freely expanding ejecta of the progenitor star and $\alpha = 3/5$ when 
expansion is in an uniform ISM. Recent measurements of the radio expansion velocity of 
3C58 (Bietenholz et al. 2001) yield a value of $\sim 900$ km/sec, which is in 
contrast with the observed values for $R_{\rm pwn} = 4.5$ pc and $t_{\rm age} = 821$ 
years (assuming an association with the historical SN event of 1181 and a power-law
expansion rate), which would suggest an expansion velocity of $\sim 5000$ km/sec.

One way out of this discrepancy is to suggest that the remnant is much older. However
this would suggest that the pulsar wind is interacting directly with the
ISM, and there was no freely expanding envelope due to a very low-energy supernova
event (Bietenholz et al. 2001). If this scenario represents the evolutionary
status of 3C58, the shape of 3C58 corresponds exactly with the shape from the 
dynamical models presented in this paper.

Another scenario for 3C58 was given by van der Swaluw et al (2001), where it is 
suggested that the PWN of 3C58 is being compressed by the reverse shock. This 
scenario can explain a number of observational properties of 3C58 and would still 
associate the remnant with the presupernova progenitor of SN 1181:

\begin{itemize}

\item
The compression of the PWN by the reverse shock would decrease the expansion 
velocity, which is too low in the case of 3C58 if it was supersonically expanding 
(Bietenholz et al. 2001).

\item
The compression of the PWN increases the magnetic field and particle energies,
this can lead to an increase of the radio flux density as is observed in 3C58 
in the period between 1967 and 1986 (Green 1987).

\item
The compression of the PWN can bring down the frequency of the synchrotron cooling 
break (Gallant et al. 2002), which is at a comparatively low frequency for 3C58
(see e.g. Woltjer et al. 1997).

\item
The passage of the reverse shock and the subsequent compression would also increase 
the pressure inside the PWN. This would decrease the radius of the wind termination 
shock $R_{\rm ts}$. The ratio between the termination shock $R_{\rm ts}$ and the boundary 
of the PWN $R_{\rm pwn}$ is indeed remarkably small in 3C58, i.e. 
$R_{\rm ts}/R_{\rm pwn} \sim 0.01$ (Frail \& Moffett 1993).

\end{itemize}

If the above scenario is correct, the absence of any detected shell corresponding with the
blastwave of the SNR remains a puzzle. 



\section{Conclusions}

I have elaborated on the work of B\&L by performing axially symmetric 2.5D MHD
simulations of a magnetized pulsar wind interacting with either a uniform interstellar 
medium or freely expanding ejecta. Both simulations were performed while the pulsar
wind bubble is bounded by a strong shock. 

The simulations only included a toroidal magnetic field component, which was injected
around the position of the pulsar. The simulations show an overall decline of the total
pressure in the radial direction in order to balance the magnetic pinching force
due to the presence of these toroidal magnetic fields. The simulations show that the PWN
is elongated as expected from the pressure differences along the PWN shock. Furthermore
the simulations show that significant elongation occurs at early stages of the expansion,
while at later stages the elongation saturates. Finally by performing 
several simulations of a pulsar wind with different values for the magnetization 
parameter, I conclude that the elongation depends weakly on the value of the 
magnetization parameter.

Although the simulations clearly show the dynamical importance of the magnetic fields,
I have shown that one can still use a simple power-law expansion rate 
($R_{\rm pwn}\propto t^\alpha$) to model the expansion of the PWN. 

I have applied the implications of the current simulations on the naked plerion 3C58,
which is an example of a PWN driven by a pulsar wind with a detected pulsar in the
center. The morphology of this plerion fits perfectly well with the simulations performed for
a magnetized pulsar wind interacting with the ISM. However this would imply that the age
of the remnant is much older and can not be associated with the historical supernova event
in 1181. Alternatively there is observational evidence that 3C58 might be an example of a PWN 
being compressed by the reverse shock of the associated SNR. This would safeguard the association
with the presupernova progenitor of SN 1181. Therefore in a future paper I will investigate the
physics of a magnetized pulsar wind embedded in an expanding SNR. In such a system the reverse shock 
will interact with an {\it elongated} PWN and would show the evolving plerionic morphology.  

\begin{acknowledgements}
I would like to thank Bram Achterberg, Luke Drury, Rony Keppens and Jan Kuijpers
for useful discussions. Furthermore I would like to thank an anonymous referee
for usefull suggestions. The Versatile Advection Code was developed as part of 
the {\em Massief Parallel Rekenen} (Massive Parallel Computing) program funded 
by the Dutch Organisation for Scientific Research (NWO). 
\end{acknowledgements}

\end{document}